\pdfoutput=1

\documentclass[11pt]{article}

\usepackage[]{EMNLP2022}

\usepackage{times}
\usepackage{latexsym}
\usepackage{multirow}
\usepackage{makecell}

\usepackage{CJKutf8}

\usepackage{amsmath} 
\usepackage{amssymb} 
\usepackage[T1]{fontenc}

\usepackage[utf8]{inputenc}

\usepackage{microtype}

\usepackage{inconsolata}
\usepackage{algpseudocode}
\usepackage{algorithm}
\usepackage{graphicx}

\algnewcommand{\Initialize}[1]{%
  \State \textbf{Initialize:}
  \Statex \hspace*{\algorithmicindent}\parbox[t]{.8\linewidth}{\raggedright #1}
}

\title{PILE: Pairwise Iterative Logits Ensemble for Multi-Teacher Labeled Distillation}

\author{Lianshang Cai\thanks{\; Equal contribution.}, Linhao Zhang\footnotemark[1], Dehong Ma\thanks{\; Corresponding authors.}, Jun Fan \\ 
        {\bf Daiting Shi, Yi Wu, Zhicong Cheng, Simiu Gu, Dawei Yin\footnotemark[2]} \\
        Baidu Inc., Beijing, China \\
        \texttt{\{cailianshang,zhanglinhao,madehong,fanjun\}@baidu.com} \\
        \texttt{\{shidaiting01,wuyi01,chengzhicong01,gusimiu\}@baidu.com} \\ 
        \texttt{yindawei@acm.org}}
        
\begin{document}

\maketitle

\begin{abstract}
Pre-trained language models have become a crucial part of ranking systems and achieved very impressive effects recently. To maintain high performance while keeping efficient computations, knowledge distillation is widely used. In this paper, we focus on two key questions in knowledge distillation for ranking models: 1) how to ensemble knowledge from multi-teacher;  2) how to utilize the label information of data in the distillation process. We propose a unified algorithm called Pairwise Iterative Logits Ensemble (PILE) to tackle these two questions simultaneously. PILE ensembles multi-teacher logits supervised by label information in an iterative way and achieved competitive performance in both offline and online experiments. The proposed method has been deployed in a real-world commercial search system. 
\end{abstract}

\section{Introduction}

Search engines have been an infrastructure in the Information Age to satisfy people's needs for querying about information. In modern search engines, multi-stage pipelines are usually employed where \textit{ranking} is usually known as the very final stage. It takes as input the shortlisted candidates of relevant documents (i.e. web pages) retrieved from previous stages, and concentrates on sorting \cite{lin2021pretrained} based on the degree of match between the latent semantics of documents and the search intent of the user's query. 

With the flourishing of pre-trained language models \cite{devlin2018bert,sun2019ernie,lan2019albert,he2020deberta}, BERT-based models have achieved state-of-the-art performance in a broad range of downstream tasks and text ranking is no exception. Pre-trained rankers based on BERT show impressive performance in ranking tasks \citep{nogueira2019passage, nogueira2019multi, zou2021pre}. 

Despite the state-of-the-art performance pre-trained models yield in laboratories, it is hardly possible to apply them directly in real-world search engines. Their large numbers of parameters go against computational efficiency while the online environment is strictly restricted in resources.
Therefore, before deploying a pre-trained model online, one necessary procedure is to reduce computational costs.

Knowledge distillation is one of the most commonly used methods to reduce the model size \citep{cristian2006model, 2015Distilling} and accelerate the computation process.  In a standard workflow of distillation, a large model (i.e. teacher) is pre-trained and finetuned well in advance, and a small model (i.e. student) imitates the teacher model's behaviors. 
The knowledge learned by the teacher model is then transferred to the student model.

One of the risk factors hindering the improvement of the student model is that the knowledge acquired by a single teacher may be insufficient and biased. 
A straightforward solution is using an ensemble of multiple teachers for knowledge transfer. The ensemble process takes the predictions of multiple teachers into account and provides comprehensive guidance that helps to improve student performance \citep{You2017LearningFM}. However, these teachers sometimes conflict with each other, and the heuristic of treating them equally by taking the mean of their predictions ignores the fact that they vary in confidence and correctness, thus often leading to suboptimal performance \citep{Du2020AgreeTD}. Besides, the valuable label information is ignored.  Hence, how to ensemble knowledge from multi-teacher and how to utilize the label information are two key questions during the distillation process.

In this work, we introduce a unified algorithm called \textbf{Pairwise Iterative Logits Ensemble (PILE)} to tackle these two questions simultaneously. 
The key idea of PILE is to assign a higher weight to teachers that produce more consistent soft targets with the golden labels. 
The resulting soft targets not only retain the generalization information transferred from the teacher models but also are integrated with the label information annotated by human experts.

We conduct both offline and online experiments and the results validate the effectiveness of PILE. The main contributions of this paper can be summarized as follows:
\begin{itemize}
  \item We propose PILE, a specially designed ensemble algorithm to tackle multi-teacher distillation and labeled distillation. To the best of our knowledge, PILE is the first work that addresses these two key questions simultaneously in the ranking scenario. 
  \item We conduct extensive offline and online experiments to demonstrate the effectiveness of our proposed method. The results show that PILE effectively boosts a real-world search engine’s performance.
\end{itemize}

\section{Related Work}
\textbf{Text Ranking}
The goal of text ranking is to generate an ordered list of texts in response to a query. Conventional learning-to-rank (LTR) techniques \cite{li2014learning} are widely used for text ranking, which plays an important role in a wide range of applications, like search engines and recommender systems. LTR techniques can be roughly categorized into three types: pointwise approach  \cite{cooper1992probabilistic,li2007mcrank}, pairwise approach \cite{joachims2002optimizing,zheng2007regression}, and listwise approach \cite{cao2007learning, burges2010ranknet}. The former two are more widely used in practice as they are easier to optimize.

Recently, deep learning approaches have been widely adopted in ranking and BERT-based ranking models achieve state-of-the-art ranking effectiveness. For example, \citet{nogueira2019passage} use BERT-large \cite{devlin2018bert} as the backbone and feed the concatenation of query and passage text to estimate the relevant scores for passage re-ranking. \citet{nogueira2019multi} formulate the ranking problem as a pointwise and pairwise classification problem and tackle them with two BERT models in a multi-stage ranking pipeline. \citet{yilmaz2019cross} aggregate sentence-level information to estimate the relevance of the documents and transfer the learned model to capture cross-domain notions of relevance.

However, the performance improvement comes at the cost of efficiency, which limits their real-world application. There are several ways to maintain high performance while keeping efficient computations for BERT-based models, such as knowledge distillation \cite{hinton2015distilling}, weight sharing \cite{lan2019albert}, pruning \cite{pasandi2020modeling, xu2020bert}, and quantization \cite{hubara2017quantized, jacob2018quantization}. In this paper, we focus on knowledge distillation which has proven a promising way to compress large models while maintaining performance.

\textbf{Knowledge Distillation}
The idea of knowledge distillation was first introduced by \citet{cristian2006model} to train small and fast models to mimic cumbersome and complex models, without much loss in performance. \citet{2015Distilling} developed this idea further by minimizing the difference between their soft target distribution. 
With the rise of the pre-training and fine-tuning paradigm, various work has later extended this idea to large-scale pre-trained models and shown impressive results on multiple NLP tasks \cite{wang2018glue, rajpurkar2018know, lai2017racelr} with a significant gain in training efficiency. 
\citet{2019DistilBERT} conducted knowledge transfer during the pre-training phase, also known as a task-agnostic way. 
\citet{2019Patient} proposed an approach to transfer knowledge between intermediate layers in the fine-tuning stage.
\citet{2020TinyBERT} additionally used attention-based distillation and hidden states-based distillation for students to imitate teachers' behaviors in intermediate layers. \citet{wang2020minilm} introduced self-attention relation-based transfer and teacher assistants \cite{mirzadeh2020improved} to further improve the performance of students.

\textbf{Ensemble Knowledge Distillation}
There is also some other work exploring the issues of multi-teacher distillation. For example, \citet{Du2020AgreeTD} adaptively ensemble knowledge distillation to ﬁnd a better optimizing direction for the student network. \citet{wu2021one} designed a co-ﬁnetuning framework to jointly finetune multiple teachers for better collaborative knowledge distillation. 
\citet{li2021dynamic} explored the influence of teacher model adoption which is promising for improving student performance.
Different from the above work, we investigate the problem of ensemble knowledge distillation in ranking tasks and use the golden label to supervise the ensemble process.

\section{Methodology}

\subsection{Ranking Task Definition}
In a search system, the ranking task aims to measure the relative order of a set of documents $D_q = \{d_i\}_{i=1}^N$ given a query $q \in Q$, where $Q$ is a set of user queries and $D_q \subset \mathbb{D}$ is a set of $q$-related documents retrieved from a large document corpus $\mathbb{D}$ \citep{liu2021pre}. The ranking model determines the order of documents by computing the relevance score $f(q, d; \theta)$ of each query-document pair $\{(q, d_i)\}_{i=1}^{N}$, where $f$ is a scoring function parameterized by $\theta$ representing the relevance of query $q$ and document $d$ . 

As regards training procedure, the ranking model is learned by minimizing the empirical loss over the training data as 
\begin{equation*}
\mathcal{L} = \sum_{q \in Q} l(Y_{D_q}, F(q, D_q)),
\end{equation*}
where $l$ is the loss function in learning to rank, e.g. pointwise loss, pairwise loss or listwise loss, and $F(q, D_q) = \{f(q, d_i)\}_{i=1}^N$ is a set of relevance scores, $Y_{D_q} = \{y_i\}_{i=1}^N$ is a set of labels. The label $y_i$ is often assigned an integer range from 0 to 4, representing the relevance of the query-document pair $(q, d_i)$.

\subsection{Knowledge Distillation}
Due to the resource constraint, the ranking model can not directly serve online in a real production environment and we use knowledge distillation (KD) to compress the model size. In a commonly used KD framework, a large teacher model $T$ is pre-trained or finetuned well in advance, and the knowledge of the teacher is transferred to a small student $S$ by minimization of the difference between them, which can be formulated as:
\begin{equation*}
\mathcal{L}_{KD} = \sum_{x \in \mathcal{D}}L(f^S(x), f^T(x)),
\end{equation*}
where $\mathcal{D}$ denotes the training dataset and $x$ is the input sample, $f^S(\cdot)$ and $f^T(\cdot)$ represent behavior measurements of teacher and student models, and $L(\cdot)$ is a loss function to measure the difference between their behaviors. The behaviors are usually represented by soft target distributions of the last layer, hidden state distributions or other deep semantic features such as self-attention distributions and embedding layer outputs \cite{2015Distilling, sun2020mobilebert, 2020TinyBERT, wang2020minilm, wang2020minilmv2}. The methods that transfer the knowledge between the internal layers are limited in generality since the teachers and students are required to have the same model structure or align with each other in the number of layers or the size of hidden layers. Based on this consideration of generality, we perform knowledge distillation on the last layer only.

\subsection{Pairwise Iterative Logits Ensemble}\label{method}
\begin{figure*}[htp]
    \centering
    \includegraphics[width=14cm]{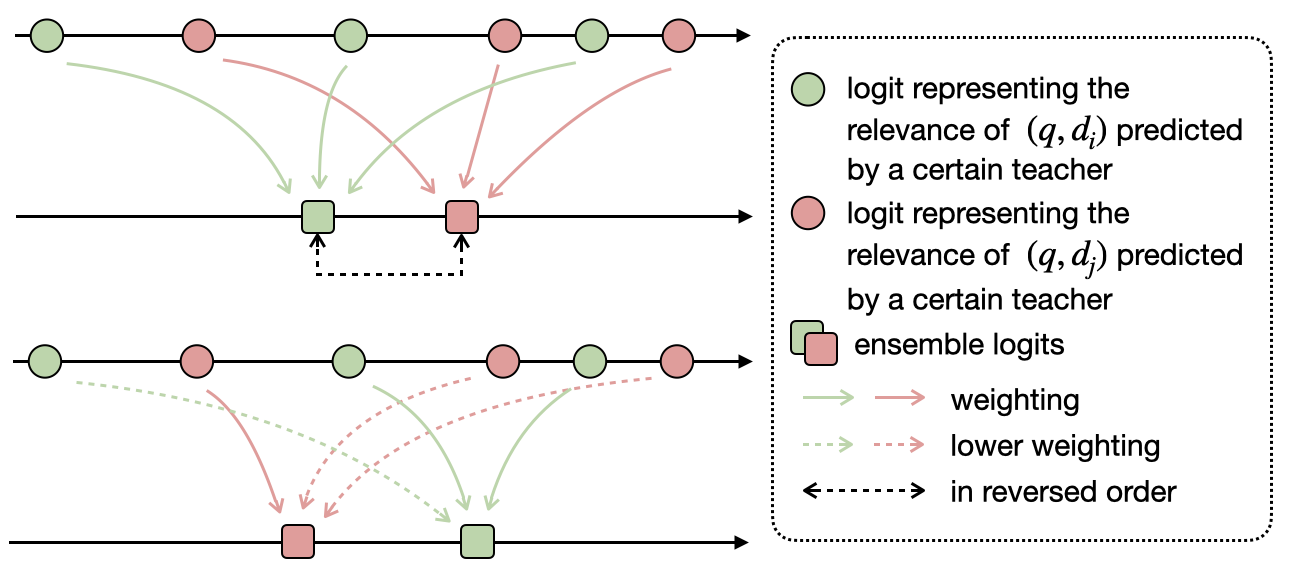}
    \caption{Procedures of PILE: 1) start with initializing the ensemble distillation logits with equal weights and 2) update a pair of resulting logits in reversed order by reassigning the weights of teachers. }
    \label{pile}
\end{figure*}
Knowledge distillation by one single teacher may bring some bias to the student model while simple yet common mitigation is distillation on the average of logits output by $n$ multiple teachers. However, the teachers with diversity may conflict with each other since the biased teacher contributes equally as the unbiased teacher, which corrodes the confidence of distillation logits. Since the logits produced by teachers on different data vary in the degree of confidence, we conduct a dynamic weighting process for the ensemble. In the ranking task, the logit predicted by the teacher for a document represents a measurement of relevance with the query. The larger logits represent more correlation between the query and the document, and the correlation information is already annotated in its label. Thus, we consider utilizing the golden label to direct the assignment of weight. The procedures of the PILE algorithm are provided in Figure \ref{pile}.

We start with initializing the ensembled distillation logit $e^{(0)}(q, d)$ for each query $q$ and document $d$ by way of averaging each teacher's outputs:

\begin{align*}
w_k(q, d) &= 1 \\
e^{(0)}(q, d) &= \frac{1}{Z(q, d)}\sum_k w_k(q, d) f_k(q, d) \\
Z(q, d) &= \sum_k w_k(q, d),
\end{align*}
where $f_k(q,d)$ represents the relevance score predicted by $k$-th teacher and $w_k(q,d)$ is its weight w.r.t query $q$ and document $d$.

Then, we perform the iterations of the update procedure. At iteration $t$, we randomly choose a pair of docs $(d_i, d_j)$ related to the same query $q$, and check whether the magnitude of their ensemble logits is consistent with their labels. More specifically, if label $y_i > y_j$ and ensemble logits $e^{(t)}(q, d_i) < e^{(t)}(q, d_j)$, we call this pair of docs in reversed order or a negative pair and consider the ensemble logits have been biased. We modify the biased ensemble logits by reassigning to zero the weight of teachers responsible for the reversed order error. The reassignment rule can be formulated as follows:
\begin{align*}
\begin{split}
 w_k(q,d_i) &= \left \{
 \begin{array}{ll}
            0, & f_k(q, d_i) < e^{(t)}(q, d_i) \\
            1, & otherwise
 \end{array}
 \right.
\end{split} \\
\begin{split}
 w_k(q,d_j) &= \left \{
 \begin{array}{ll}
            0, & f_k(q, d_j) > e^{(t)}(q, d_j) \\
            1, & otherwise
 \end{array}
 \right.
\end{split}
\end{align*}
where we assume the docs pair $(d_i, d_j)$ is in reversed order and label $y_i > y_j$ for ease of explanation. Then, we update the ensemble logits with an update rate $\lambda$:
\begin{align*}
\tilde{e}^{(t+1)}(q, d) &= \frac{1}{Z(q,d)} \sum_i w_k(q,d) f_k(q, d) \\
Z(q,d) &= \sum_k w_k(q,d) \\
e^{(t+1)}(q, d) &= (1 - \lambda) e^{(t)}(q, d) + \lambda \tilde{e}^{(t+1)}(q, d)
\end{align*}
We repeat the updating in an iterative process until the magnitude of the ensemble logits of each pair of docs is consistent with their labels or it reaches the maximum iteration number. We use the final ensemble logits to perform knowledge distillation for the student model.


\section{Experiments}
To investigate the effectiveness of our proposed method, we conduct offline experiments with baseline models and deploy our proposed model in the real-world production environment. 
In this section, we report the details of the experiment setups, datasets we used, evaluation metrics, the results of the experiments, and the case study. 

\subsection{Datasets}\label{section_datasets}
The datasets on which we pre-train, finetune, and evaluate our proposed method are collected from the Baidu search engine. For the pre-training stage, we collect a large-scale unlabeled dataset (log data) by means of the anonymous search log. The dataset contains $2,970,692,361$ query-document pairs. 

As regards finetuning stage, queries and documents are collected from the search pipelines and manually labeled on Baidu's crowdsourcing platform, where a group of hired annotators assigned an integer label range from $0$ to $4$ to each query-document pair, representing their relevance as \{bad, fair, good, excellent, perfect\}. 
We repeat the same process for the test set.
The dataset information is summarized in Table~\ref{datasets}.

\begin{table}
\centering
\begin{tabular}{c|cc}
\hline\hline
Data & \#Query & \#Query-Doc Pair\\
\hline
log data & 635,420,390 & 2,970,692,361 \\

train data & 432,410 & 8,794,863 \\
test data  & 12,044 & 289,835 \\ 
\hline\hline
\end{tabular}
\caption{\label{datasets}
Dataset statistics
}
\end{table}

\subsection{Training details}
We use a 12-layer Transformer \citep{vaswani2017attention} structure with 768 hidden sizes and 12 attention heads as the backbone of teacher models and a 6-layer Transformer structure with 768 hidden sizes and 12 attention heads as student models, where the parameters are randomly initialized, pre-trained and finetuned on the datasets described in section ~\ref{section_datasets}. 

In order to obtain multiple teachers with similar performance and some differences for ensemble knowledge distillation, we use the same pre-trained checkpoint and finetune it on different samplings of the total train data. 
Specifically, the teacher $T_1$ is finetuned on the whole train data, and teacher $T_2$ and $T_3$ are finetuned on $80\%$ of the whole train data. 
The intuition behind this is that we want all the teachers to perform relatively well but not behave as the same one otherwise the ensemble of three teachers may degenerate into a single model, which will compromise the benefits of the ensemble knowledge distillation. 
As a result, $T_1$ performs best on the test set and the other two teachers have a competitive performance as $T_1$. The results of three teachers on the test set are shown in Table~\ref{offline-comparison}. 

In the training procedure, we use the Adam optimizer \citep{kingma2014adam} with $\beta_1=0.9$ and $\beta_2=0.99$. For both 12-layer and 6-layer models, we set the learning rate as 2e-5, the batch size as 64, and the warm-up step as 1000. 
In the PILE procedure, the maximum number of iterations is set to $|D_q|^{\frac{3}{2}}$ where $|D_q|$ is the size of documents set $D_q$ related to a given query $q$ and the update rate $\lambda$ is set to $0.9$. 
In the knowledge distillation stage, we set the warm-up step as $1000$, the learning rate as 2e-5 and the batch size as 1024.

\subsection{Evaluation Metrics}
The evaluation metrics we used in the experiments are as follows.

The \textbf{Positive-Negative Ratio (PNR)} measures the consistency between the golden labels and the scores output by models. For a given query $q$ and a list of $N$ associated documents ranked by model, the PNR can be calculated by this formulation:

\begin{small}
\begin{equation*}
PNR = \frac{\sum_{i,j\in[1,N]}I\{y_i>y_j\}I\{f(q,d_i)>f(q,d_j)\}}{\sum_{i,j\in[1,N]}I\{y_i>y_j\}I\{f(q,d_i)<f(q,d_j)\}},
\end{equation*}
\end{small}
where $I\{\cdot\}$ is the indicator function, taking the value $1$ if the internal statement is true or $0$ otherwise. For the test set that contains a good many queries, we average PNR values over all queries.

The \textbf{Discounted Cumulative Gain (DCG)} \citep{Jrvelin2000IREM} is a widely used metric that evaluates the ranking result of search engines. More specifically, DCG is calculated as a weighted sum of the document's relevance degree $G_i$ at each position $i$, where the weight is assigned according to the document's position in the ranking results:
\begin{equation*}
DCG = \sum_i\frac{G_i}{log_2(i + 1)}
\end{equation*}

The \textbf{Interleaving} \citep{Chapelle2012LargescaleVA, 10.1145/2668120} is extensively used for comparing a new system with the base system in industrial information systems evaluation. The results of two systems are interleaved and presented together to the end users, whose clicks would be credited to the system that provides the corresponding results. The gain of the new system $A$ over the base system $B$ can be denoted as $\Delta_{AB}$:
\begin{equation*}
\Delta_{AB} = 0.5 * \frac{wins(A) - wins(B)}{wins(A) + wins(B) + ties(A, B)}, 
\end{equation*}
where $wins(A)$(or $wins(B)$) counts the number of times when the results produced by system $A$ (or $B$) are more preferred than the other system for a given query and $ties(A, B)$ counts the number of times when the two systems are tied.

We also conduct a comparison called \textbf{Good vs. Same vs. Bad (GSB)} between two systems by inviting professional annotators to estimate which system produced a greater ranking result for each given query \citep{Zhao2011AutomaticallyGQ}. The gain of a new system can be formulated as:
\begin{equation*}
\Delta_{GSB} = \frac{\#Good - \#Bad}{\#Good + \#Bad + \#Same},
\end{equation*}
where $\#Good$ (or $\#Bad$) denotes the number of queries that the new (or base) system provides better ranking results and $\#Same$ for the number of results that are equal in quality.

\subsection{Offline Experimental Results}
We conduct several comparison experiments to verify our proposed method. The models in the offline comparison experiments include: 
\begin{itemize}
    \item  \textbf{Base}: We use an ERNIE-based ranking model as our base model, which is finetuned with a pairwise loss using human-labeled query-document pairs without any guidelines from teachers; 
    \item  \textbf{single-KD}: In this setting we add knowledge distillation loss when training the base model using the teacher that performs best on the test set;
    \item  \textbf{AE-KD}: Instead of using the single teacher, this variant uses an ensemble of 3 teachers with averaged weight to perform knowledge distillation;
    \item  \textbf{PILE-KD}: When performing knowledge distillation, PILE-KD uses human-annotated labels with the help of the PILE algorithm to conduct a dynamic weighting process for the ensemble of 3 teachers.
\end{itemize}

\begin{table}
\centering
\begin{tabular}{lcc}
\hline\hline
Method & PNR & Improvement \\
\hline
Teacher1 & 3.21 & - \\
Teacher2 & 3.20 & - \\
Teacher3 & 3.19 & - \\
\hline
Base & 3.11 & - \\
+ single-KD & 3.15 & +1.29\% \\
+ AE-KD   & 3.16 & +1.61\% \\
+ PILE-KD   & \textbf{3.18} & +2.25\% \\
\hline\hline
\end{tabular}
\caption{\label{offline-comparison}
Offline comparison of the proposed methods. 
}
\end{table}

The results of each model are shown in Table~\ref{offline-comparison} with the improvement compared to the base model. We also report the performance of the teachers used in knowledge distillation.
As we expected, all the distilled models consistently outperform finetuned base model thanks to teacher models' guidance and regularization. And besides, using an ensemble of teachers gains further promotion than the single teacher distillation. After ensembling multiple teachers by averaging distillation logits, the PNR reaches 3.16, exceeding the base by 1.61\%. This shows that the remission from biased distillation by the cooperation of multiple teachers improves students in semantic matching. Moreover, by applying the PILE algorithm, we can see that the student can beat the base model by a large margin w.r.t PNR, where the value is improved to 3.18 by 2.25\% improvement. It shows the effectiveness of dynamic reduction of biased teachers' weight in the ensemble process.

\subsection{Online Experimental Results}

\begin{table}
\centering
\begin{tabular}{c|cc}
\hline\hline
 &Random & Tail\\
\hline
$\Delta$DCG & +0.10\% & +0.27\% \\
$\Delta$GSB & +3.70\% & +1.62\%\\
\hline\hline
\end{tabular}
\\[10pt]
\setlength{\tabcolsep}{1mm}{
\begin{tabular}{c|cc|cc}
\hline\hline
\multirow{2}{1em}{ } &\multicolumn{2}{|c|}{Query Type} &\multicolumn{2}{|c}{Query Length} \\
                     & Random   & Tail      & short   & long \\
\hline
$\Delta_{AB}$        & +0.022\% & +0.029\%  & +0.01\% & +0.039\% \\
\hline\hline
\end{tabular}}
\caption{\label{online-comparison}
Online comparison of the proposed methods.
}
\end{table}

To investigate the effectiveness of our proposed method in the real production environment, we deploy the proposed model in Baidu Search, a widely used Chinese search engine, and conduct online experiments for comparison.

The results are presented in Table~\ref{online-comparison}, which comprises the performance comparison regarding $\Delta DCG$, $\Delta GSB$, and $\Delta_{AB}$. 
We consider the queries from the perspective of types and lengths in the search log. The tail queries and long queries are the queries whose search frequency is lower than 10 times per week or whose length is greater than 10 respectively. 
Since the heterogeneous search queries follow long-tail distributions, the tail queries make up a significant part of the queries in the search engine. As we can see the proposed method improves the performance of the online ranking system consistently. Particularly, we can observe that the gains of tail queries in the $\Delta DCG$ and $\Delta GSB$ for our method are 0.27\% and 1.62\% respectively. 
Compared with AE-KD, PILE-KD can enable students to retain the ability of teachers as much as possible, and the improvement on long-tail queries also confirms this.

\subsection{Ablation Studies}
To illustrate the detailed effects of the proposed algorithm, we take a deep insight into the contributions   of each setting. 

\begin{figure}
    \centering
    \includegraphics[width=7.5cm]{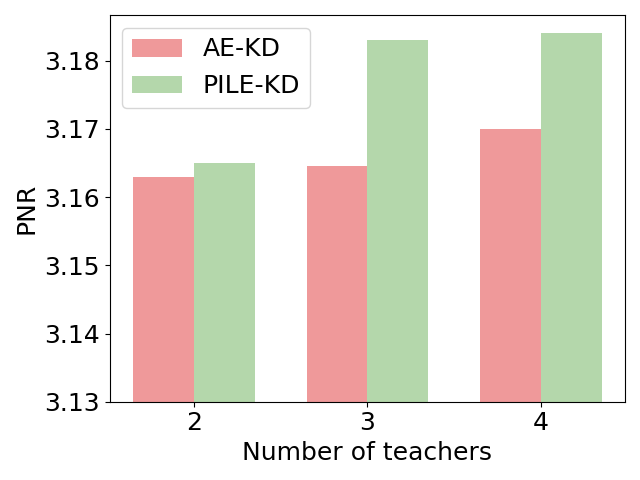}
    \caption{The effect of the number of teachers.}
    \label{num-teacher}
\end{figure}

\begin{figure}
    \centering
    \includegraphics[width=7.5cm]{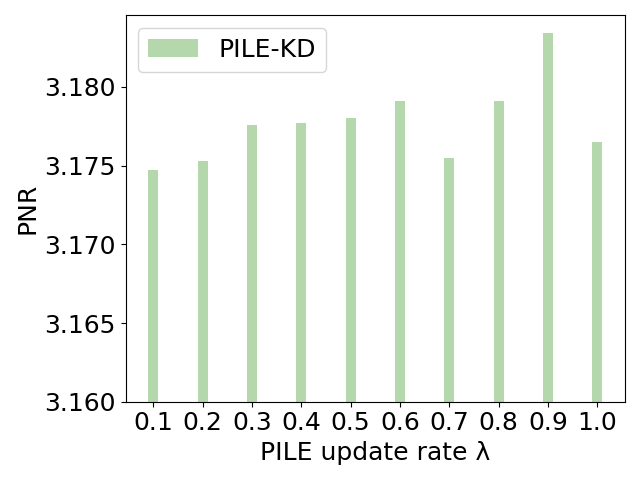}
    \caption{The effect of update rate $\lambda$ in PILE.}
    \label{update-rate}
\end{figure}

\textbf{Number of teachers.}
We first focus on the effect of the number of teachers and the results are shown in
Figure\ref{num-teacher}. As expected, the PNRs under both AE-KD and PILE-KD settings increase with the number of teachers and consistently outperform the result of the single-KD model, showing the benefits of using multiple teachers for distillation. Besides, the performance under PILE-KD settings has always been better than it under AE-KD settings, showing the benefits of reducing noise and bias with the help of the label information annotated by human experts in the process of ensemble knowledge distillation.

\textbf{PILE update rate.}
We further examine the effect of the PILE update rate $\lambda$. As shown in section \ref{method}, the update rate $\lambda$ controls the smoothness of two successive iterations. As we can see in Figure \ref{update-rate}, the PNR result increases with the $\lambda$ ranging from 0.1 to 1.0, until reaching the apex at 0.9. The larger update rate $\lambda$ makes the reliable teachers get more weight in a PILE iteration, resulting in more dependable soft targets. The results further prove that the distillation process can benefit from PILE iteration.

\subsection{Case Study}

To illustrate the effect of the PILE algorithm in the ensemble knowledge distillation concretely, we show a case that corrects the ensemble logits of two documents that are in reversed order. 

As we can see in Table~\ref{case_study}, two documents that are related to the same query are labeled by annotators as $0$ and $3$ respectively, representing their relevance with the query. The teachers' predictions of the relevance are listed in the table as $T_1\sim T_3$. In the last two rows of the table, we show the ensemble results using averaged weight (AE) and the PILE method respectively.

\begin{CJK*}{UTF8}{gbsn}
\begin{table}[h]
\centering
\begin{tabular}{p{0.6cm}|p{3cm}|p{3cm}}
\hline\hline
query &  \multicolumn{2}{p{6.5cm}}{北京驾校教练一个月能挣多少钱？(How much does a Beijing driving school coach earn a month?)} \\
\hline
doc & 北京私人教练一个月能挣多少钱？(How much does a Beijing personal trainer make a month?) & 驾校教练工资一月多少？百度知道\ (How much is the salary of a driving school coach per month? Baidu Knows)\\
\hline
label    & 0      & 3      \\
\hline\hline
$T_1$ & 0.0589 & 0.0271\\
\hline
$T_2$ & 0.1923 & 0.0331 \\
\hline
$T_3$ & 0.1057 & 0.0983\\
\hline
AE       & 0.1190 & 0.0528 \\
\hline
PILE     & 0.0590 & 0.0981 \\

\hline\hline
\end{tabular}

\caption{\label{case_study}
The teachers' logits and ensemble logits for two documents.
}
\end{table}
\end{CJK*}

To get the ensemble logits for knowledge distillation, AE takes the mean of teachers' predictions. However, influenced by individual teachers, the result ensemble logits of the two documents are contrary to their golden labels. In other words, the document with higher relevance is scored lower than the irrelevant one after the ensemble process, which will confuse the student in the knowledge transfer process. Benefiting from the PILE iteration, the teachers consistent with the golden label are assigned more weight. The resulting soft targets not only retain the knowledge that transfers from teachers but also are integrated with the label information annotated by human experts, which is more promising for knowledge distillation.

\section{Conclusion}
In this work, we propose an easy-to-implement approach to multi-teacher distillation for large-scale ranking models. Our algorithm ensembles multi-teacher logits supervised by human-annotated labels in an iterative way. We conduct the offline experiments as well as deploy our methods in an online commercial search system which demonstrates its superiority.

\section*{Acknowledgements}

The authors would like to thank the colleagues at Baidu Inc. for their constructive suggestions that helped improve the paper, and hope everything goes well with their work. The authors are also indebted to the anonymous reviewers for their valuable comments and suggestions on the paper.



\label{sec:appendix}

\bibliography{anthology,custom}
\bibliographystyle{acl_natbib}
\end{document}